\documentclass[letterpaper]{JHEP3}

\usepackage{amsthm}
\usepackage{amscd}
\usepackage{epsfig}

\newcommand{\bastar}{\begin{eqnarray*}}
\newcommand{\eastar}{\end{eqnarray*}}
\newskip\humongous \humongous=0pt plus 1000pt minus 1000pt

\newif\ifdtup

\relax
\newcommand{\be}{\begin{equation}}
\newcommand{\ee}{\end{equation}}
\newcommand{\bea}{\begin{eqnarray}}
\newcommand{\eea}{\end{eqnarray}}
\newcommand{\X}{{\vec X}}
\newcommand{\pro}{\partial}
\newcommand{\n}{\hat n}
\newcommand{\oneg}{\displaystyle\frac{1}{g}}

\newcommand{\D}{{\hat D}}

\newcommand{\A}{{\vec A}}
\newcommand{\valpha}{{\vec \alpha}}

\newcommand{\hn}{{\hat n}}
\newcommand{\hD}{{\hat D}}
\newcommand{\dfrac}{\displaystyle\frac}
\newcommand{\ba}{\begin{array}}
\newcommand{\ea}{\end{array}}

\newcommand{\nn}{\nonumber}

\newcommand{\bD}{\bar D}
\title{Color Reflection Invariance and Monopole Condensation in QCD}

\author{Y. M. Cho\\
C.N.Yang Institute for Theoretical Physics \\
State University of New York, Stony Brook \\ 
New York 11794, USA \\
and \\
School of Physics \\ 
College of Natural Sciences, Seoul National University \\ 
Seoul 151-747, Korea\\
E-mail: \email{ymcho@yongmin.snu.ac.kr}}
\abstract{We review the
quantum instability of the Savvidy-Nielsen-Olesen (SNO) vacuum of
the one-loop effective action of $SU(2)$ QCD, and point out a
critical defect in the calculation of the functional determinant
of the gluon loop in the SNO effective action. We prove that the
gauge invariance, in particular the color reflection invariance,
exclude the unstable tachyonic modes from the gluon loop integral.
This guarantees the stability of the magnetic condensation in
QCD.}
\preprint

\begin{document}

\section{Introduction}

The confinement problem in quantum chromodynamics (QCD)
is probably one of the most challenging problems
in theoretical physics. It has
long been argued that the confinement in QCD
can be triggered by the monopole condensation \cite{nambu,cho1}.
Indeed, if one assumes monopole condensation, one can easily argue
that the ensuing dual Meissner effect could guarantee the confinement
of color \cite{nambu,cho1,cho2}. But it has been extremely difficult
to prove the monopole condensation in QCD. Although
the monopole condensation has been established in a supersymmetric
QCD \cite{witt}, a satisfactory theoretical proof of the desired
monopole condensation in the conventional QCD has remained
elusive.

A natural way to establish the monopole condensation in QCD
is to demonstrate that the quantum fluctuation triggers
a phase transition through the dimensional transmutation
known as the Coleman-Weinberg mechanism \cite{cole},
by generating a non-trivial vacuum which can be identified
as a monopole condensation. Coleman and Weinberg have demonstrated that
the quantum effect could trigger a phase transition
in massless scalar QED with a quartic self-interaction of
scalar electrons, by showing that the one-loop effective action generates
a condensation of scalar electrons which defines a non-trivial new vacuum.
To prove the monopole condensation, one need to demonstrate
such a phase transition in QCD.
There have been many attempts to do so
with the one-loop effective action of QCD using the background
field method \cite{savv,niel,ditt}. Savvidy has first calculated
the effective action of $SU(2)$ QCD in the presence of an {\it ad hoc}
color magnetic background, and has almost ``proved''
the magnetic condensation in QCD. In particular, he
showed that the quantum effective potential obtained
from the real part of the one-loop effective action
has the minimum at a non-vanishing magnetic background \cite{savv}.
This is exactly what everybody was looking for.
Unfortunately, this calculation was repeated by Nielsen and
Olesen, who showed that the magnetic background
generates an extra imaginary part in the effective action
which induces the pair-creation of the gluons and thus
destablizes the magnetic condensation \cite{niel}.
This instability of the ``Savvidy-Nielsen-Olesen (SNO)
vacuum'' has never been seriously
challenged, and destroyed the hope to establish the monopole condensation
in QCD with the effective action \cite{ditt}.

A few years ago, however, there has been a new attempt to calculate
the one-loop effective action of QCD
with a gauge independent separation
of the non-Abelian monopole background from
the quantum field \cite{cho3}. Remarkably, in this
calculation the effective action has been shown to
produce no imaginary part in the presence of
the monopole background, but a negative imaginary part
in the presence
of the pure color electric background. This implies that in QCD
the non-Abelian monopole background produces a stable monopole
condensation, but the color electric background
becomes unstable by generating a pair annhilation of
the valence gluon at one-loop level.
The new result sharply contradicts with the earlier results,
in particular on the stability of the monopole condensation.
This has resurrected the old controversy on the stability of monopole
condensation.

To resolve the controversy it is important to understand
the origin of the instability of the SNO vacuum.
It is well-known that the energy of a charged vector field
moving around a constant magnetic field depends on the
spin orientation of the vector field, and when the
spin is anti-parallel to the magnetic field, the zeroth
Landau level has a negative energy.
Because of this the functional determinant of the gluon loop
in the SNO magnetic background necessarily contains
negative eigenvalues which create a severe infra-red divergence
in the effective action \cite{niel}. And, when one regularizes
this divergence with the $\zeta$-function regularization,
one obtains the well-known imaginary
component in the effective action which destablizes
the magnetic condensation.
This shows that the instability of the SNO vacuum originates
from the the negative eigenvalues of the functional determinant.
Since the existence of the negative eigenvalues is so obvious,
the instability of the SNO vacuum has become the prevailing
view which nobody has dared to challenge \cite{niel,ditt}.

This popular view, however, is not without defect.
To see this notice that the eigenfuctions corresponding to
the negative eigenvalues describes the tachyons which violate
the causality and thus become unphysical. This implies that
one should exclude these tachyons
in the calculation of the effective action.
But the standard $\zeta$-function regularization fails
to remove the contribution of the tachyonic eigenstates
because it is insensitive to causality. On the other hand,
if we adopt the infra-red regularization which respects the
causality, the resulting effective action no longer has
the imaginary part \cite{cho3}. But since the $\zeta$-function
regularization has worked so
well in quantum field theory, there seems no compelling reason
why it should not work in QCD. So we need an independent
argument which can support the stability of the magnetic
condensation in QCD.

One way to check the stability of the magnetic condensation
is to calculate the imaginary part of the one-loop effective action
with a perturbative method. This idea was first proposed
by Schanbacher, but has never been taken seriously
till recently \cite{sch}.
This is understandable because the monopole condesnation is
supposed to be a non-perturbative
effect, and it is highly unlikely that one can calculate
a non-perturbative effect with a perturbative method.
The massless gauge theories, however,
have a very unique feature that the imaginary part of
the one-loop effective action is propotional to $g^2$, where $g$
is the coupling constant. This is true in
both QCD \cite{niel,ditt} and massless QED
\cite{cho01,cho5}. This enables us to calculate the imaginary part of
the effective action with a perturbative method.
Remarkably the recent perturbative calculation has
confirmed that the effective action
should have no imaginary part in the magnetic background \cite{cho4}.
This might sound surprising but could really have been expected,
because the perturbative calculation is based on the causality
which naturally excludes the contribution of tachyonic eigenstates
in the calculation of the effective action.

The perturbative confirmation of the infra-red regularization
by causality should settle the controversy on the stability of the
monopole condensation. But this does not settle the controversy
completely. There are more questions to be answered.
The perturbative calculation does tell
that the tachyonic modes should be
excluded in the calculation of the effective action,
because they violate the causality. If so, they should have
been excluded in the calculation of the functional determinant.
Unfortunately the perturbative calculation does not
tell exactly what went wrong in the earlier calculation of
the SNO effective action. In particular, it does not tell why one
could not exclude the tachyonic modes in the calculation of
the functional determinant. Considering the fact that
the monopole condensation is such an important issue for
the confinement in QCD, one can not easily dismiss the
instability of the SNO vacuum before one figures out exactly how
one can remove them from the functional determinant. Since both
the infra-red regularization by causality
and the perturbative calculation are based on the causality
principle, it would be more convincing if one could show
this with an independent principle.

Fortunately we do have an independent principle, the principle
of gauge invariance, which allows us to demonstrate this.
The SNO vacuum is not gauge invariant, and the instability of
the SNO vacuum has been attributed to this defect.
To cure this defect Nielsen and Olesen have introduced
``the Copenhagen vacuum'' which is made of gauge invariant
combination of blockwise randomly oriented color magnetic fields,
and suggested that such a gauge invariant vacuum could
generate a stable magnetic condensation \cite{niel}.
But this Copenhagen vacuum, although conceptually appealing,
has not been so useful to prove the monopole condensation.

{\it The purpose of this paper is to show that a proper
implementation of the gauge invariance in the calculation of
the functional determinant of the gluon loop excludes
the unstable tachyonic modes,
and thus naturally restore the stability of the magnetic background.
This suggests that it is
the incorrect calculation of the functional determinant,
not the $\zeta$-function regularization, which causes
the instability of the SNO vacuum}.
This means that tachyons should not have been
there in the first place. They were there to create
a mirage, not physical states. In the absence of
the tachyons, of course, there is no instability of the SNO vacuum.
This vindicates the $\zeta$-function regularization.
It is simply too honest to correct the incorrect calculation of
the functional determinant.

In the old approach Savvidy starts from an {\it ad hoc} magnetic
background which is not gauge invariant \cite{savv,niel}.
Because of this the functional determinant of the gluon loop
contains the tachyonic eigenstates when the gluon spin is
anti-parallel to the magnetic field, which in turn
develops an imaginary part in the effective action
and destabilizes the SNO vacuum. In the following, however,
we show that the spin polarization of the gluon
is not a gauge independent concept. The reason is that one can
change the direction of the magnetic field with a simple gauge
transformation, so that the spin flip of the gluon
corresponds to a gauge transformation. This means that
the gauge invariant part of the functional determinant
should not contain any negative eigenvalue.
This tells that, if we impose the gauge invariance properly,
the instability of the SNO background should disappear.

In our approach we start from a gauge invariant
non-Abelian monopole background from the beginning \cite{cho3,cho4}.
This precludes the tachyonic eigenstates
to enter in the calculation of the effective action.
In this paper we show that a natural way
to make the monopole background gauge invariant is to impose the color
reflection invariance to the vacuum, and show how this color
reflection invariance removes the contribution
of tachyonic modes in the functional determinant of the gluon loop.
In fact we show that this gauge invariant calculation produces
exactly the same effective action we obtain with
the infra-red regularization by causality.

The paper is organized as follows. In Section II
we review the background field method to calculate
the quantum effective action of QCD. In Section III
we rederive the old SNO effective action,
and discuss how the $\zeta$-function regularization creates
the instability of the SNO vacuum. In Section IV we discuss
the gauge independent separation of the monopole background
from the quantum fluctuation, and compare
the monopole background with the gauge dependent SNO background.
In Section V we review the infra-red regularization by causality,
and show how the infra-red regularization by causality generates a stable
monopole condensation in QCD. In Section VI we briefly
discuss the perturbative calculation of the imaginary part of
the QCD effective action, and show how the perturbative calculation
endorses the the infra-red regularization by causality.
In Section VII we review the color reflection invarince in QCD,
and show how the reflection invarince excludes the tachyonic modes
from the functional determinant of the gluon loop and assures
the stability of the monopole condensation. In Section VIII we discuss
the physical meaning of our analysis, in particular
the color reflection invarince, in connection with
the confinement in QCD.

\section{Background Field Method: A Review}

In this section we review the background field method \cite{dewitt,pesk}
to obtain the one-loop effective action of QCD, and
derive the integral expression of the QCD effective action.
For simplicity we will concentrate on $SU(2)$ QCD in this paper.

To obtain the one-loop effective action one must first divide
the gauge potential $\vec A_\mu$ into two parts, the slow-varying
classical background $\vec B_\mu$ and the fluctuating quantum
part $\vec Q_\mu$,
\bea
\vec A_\mu = \vec B_\mu + \vec Q_\mu,
\label{d}
\eea
and integrate out the quantum part
with a functional integration. To do this remember that
(\ref{d}) allows two types of gauge transformations, the background
gauge transformation described by
\bea
&\delta \vec B_\mu = \dfrac{1}{g} \bar D_\mu \vec \alpha,
~~~~~\delta \vec Q_\mu = - \vec \alpha \times \vec Q_\mu, \nn\\
&\bar D_\mu = \partial_\mu + g \vec B_\mu \times,
\label{bgt}
\eea
and the physical gauge transformation described by
\bea
\delta \vec B_\mu = 0,
~~~~~\delta \vec Q_\mu = \dfrac{1}{g} D_\mu \vec \alpha,
\label{pgt}
\eea
where $\vec \alpha$ is an infinitesimal gauge parameter.
Notice that both (\ref{bgt}) and (\ref{pgt}) satisfy
\bea
\delta \vec A_\mu = \dfrac{1}{g} D_\mu \vec \alpha.
\label{gt}
\eea
To integrate out the quantum field one may impose
the following gauge condition to the quantum fields,
\bea
&\vec F = \bar D_\mu \vec Q_\mu = 0, \nn\\
&{\cal L}_{gf}=- \dfrac{1}{2\xi} (\bar D_\mu \vec Q_\mu)^2.
\label{gc}
\eea
The corresponding Faddeev-Popov determinant is given by
\bea
M^{ab}_{FP} = \dfrac {\delta F^a}{\delta \alpha^b} = (\bar D_\mu D_\mu)^{ab}.
\label{fpd}
\eea
With this gauge fixing
the effective action takes the following form,
\bea
&\exp ~\Big[iS_{eff}(\vec B_\mu) \Big] = \dfrac{}{} \int {\cal D}
\vec Q_\mu {\cal D} \vec{c} ~{\cal D}\vec{c}^{~*}
\exp \Big\{~i \dfrac{}{} \int \Big[-\dfrac {1}{4} \vec G_{\mu\nu}^2
-\dfrac{1}{4} (\bar D_\mu \vec Q_\nu
- \bar D_\nu \vec Q_\mu)^2 \nn\\
&-\dfrac{g}{2} \vec G_{\mu\nu} \cdot (\vec Q_\mu \times \vec Q_\nu)
-\dfrac{g^2}{4}(\vec Q_\mu \times \vec Q_\nu)^2
+\vec{c}^{~*}\bar {D}_\mu D_\mu\vec{c}
-\frac{1}{2\xi} (\bar D_\mu \vec Q_\mu)^2 \Big] d^4x \Big\},
\label{ea}
\eea
where $\vec c$ and ${\vec c}^{~*}$ are the ghost fields.
Notice that the effective action
(\ref{ea}) is explicitly invariant under the background
gauge transformation (\ref{bgt}) which involves
only $\vec B_\mu$, if we add the following
gauge transformation of the ghost fields to (\ref{bgt}),
\bea
\delta \vec c = - \alpha \times \vec c,
~~~~~\delta \vec c^{~*} = - \alpha \times \vec c^{~*}.
\eea
This guarantees the gauge invariance of the resulting
effective action.

The gluon loop and the ghost loop integrals
give the following functional deteminants (at one-loop level)
\bea
&{\rm Det}^{-\frac{1}{2}} K_{\mu \nu}^{ab} =
{\rm Det}^{-\frac{1}{2}} \Big(-g_{\mu \nu}
\bD^2_{ab}- 2g \epsilon_{abc} G_{\mu \nu}^c \Big),\nn \\
& {\rm Det} M_{ab} = {\rm Det} \Big(-\bD^2_{ab} \Big),
\label{fd}
\eea
from which one has
\bea
\Delta S = \dfrac{i}{2} \ln {\rm Det} K - i \ln {\rm Det} M.
\label{ea0}
\eea
So the correct calculation of the determinants becomes crucial
for us to obtain the effective action.

\section{SNO Effective Action: A Review}

Savvidy, and Nielsen and Olesen have chosen a covariantly constant
color magnetic field as the classical background
in their calculation of the effective action \cite{savv,niel,ditt}
\bea
&\vec B_\mu = \bar B_\mu \n_0 = \dfrac{1}{2} \bar H_{\mu\nu} x_\nu \n_0,
~~~~~\vec G_{\mu\nu} = \bar H_{\mu\nu} \n_0, \nn\\
&\bar D_\mu \vec G_{\mu\nu} = 0,
\label{sb}
\eea
where $\bar H_{\mu\nu}$ is a constant magnetic field and $\n_0$ is
a constant unit isovector ($\n_0^2=1$).
With the background, one can calculate the functional
determinant (\ref{fd}).
The calculation of the determinant amounts to the calculation of
the eigenvalues of the determinant. Nielsen and Olesen have
pointed out that
this reduces to the calculation of the energy eigenvalues of
a massless charged vector field $X_\mu$ in a constant external
magnetic field $\bar H_{\mu\nu}$ \cite{niel},
\bea
&E^2 X_\mu = [-\bar D^2 g_{\mu\nu} + \bar D_\mu \bar D_\nu
+ 2ig \bar H_{\mu\nu}] X_\nu =0, \nn\\
& \bar D_\mu = \partial_\mu + ig \bar B_\mu. \label{eveh}
\eea
Suppose the magnetic field is
in $z$-direction. Then this has the well-known eigenvalues
\bea
&E^2 = 2gH (n + \dfrac{1}{2}) + k^2 \pm 2gH, \nn\\
& H = \bar H_{12},
\label{ev}
\eea
where $k$ is the momentum of the eigen-function in
$z$-direction (the direction of the background magnetic field).
Notice that the $\pm$ signature correspond to the spin $S_3=\pm 1$
of the charged vector field (in the direction of the magnetic field).
So, when $n=0$,
the eigen-functions with $S_3=-1$ have an imaginary energy when
$k^2<gH$, and thus becomes tachyons which violate
the causality. The eigenvalues of the functional determinant
is shown in Fig. 1.

\begin{figure*}
\begin{center}
\includegraphics{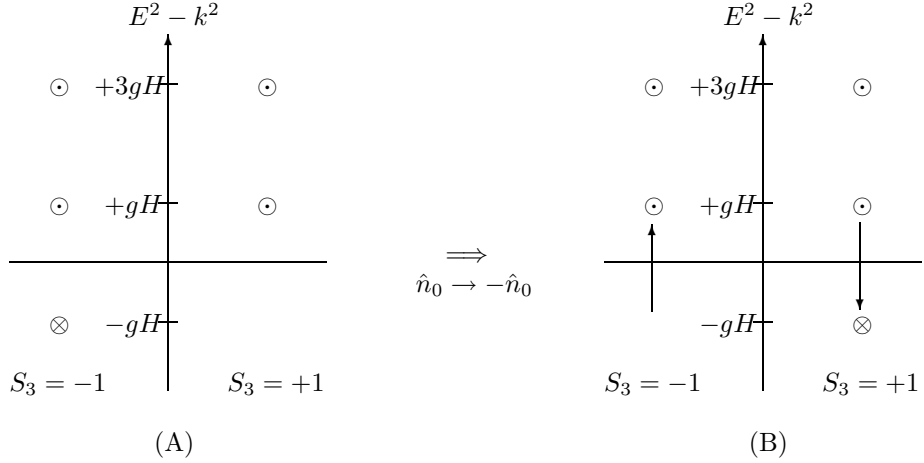}
\end{center}
\caption{\small{\label{Fig. 1} The eigenvalues of the functional
determinant of the gluon loop. When the gluon spin is
anti-parallel to the magnetic field ($S_3=-1$), the ground state
(with $n=0$) becomes tachyonic when $k^2<gH$. Notice, however,
that under the color reflection of $\hn_0$ to $-\hn_0$, the
magnetic field changes its direction while the gluon spin remains
the same. This means that the spin polarization direction of gluon
with respect to the magnetic field is a gauge artifact, which
tells that the gauge invariant functional determinant should not
include the tachyonic state.}}
\end{figure*}

An important point here is that $\n_0$ can be
rotated to $-\n_0$ by a gauge transformation. This means that
one can change the direction of the magnetic field
by a gauge transformation. And obviously the gauge transformation
does not affect the gluon spin. This means that one can change
the spin polarization direction of
the gluon with respect to the magnetic field by
a gauge transformation. But notice that the eigenstates
with $S_3=-1$ changes to the eigenstates with $S_3=+1$
(and vise versa) under the gauge transformation.
This tells that the eigenstates with $S_3=-1$ and $n=0$
are not invariant under the gauge transformation.
This point will become very important
when we make a gauge invariant calculation of
the functional determinant in Section VII.

From (\ref{ev}) one obtains
\bea
&\Delta S = i\ln {\rm Det} \Big[(-\bD^2+2gH)(-\bD^2-2gH) \Big],
\label{fdhx}
\eea
and the integral expression of the effective action
\bea
&\Delta{\cal L} = \dfrac{}{} \lim_{\epsilon \rightarrow 0}
\dfrac{1}{16 \pi^2}\int_{0}^{\infty}
\dfrac{dt}{t^{2-\epsilon}} \dfrac{gH \mu^2}{\sinh (gHt/\mu^2)} \nn\\
&\times \Big[\exp (-2gHt/\mu^2 )
+ \exp (+2gHt/\mu^2) \Big],
\label{eahx}
\eea
where $\mu^2$ is a dimensional parameter.

The effective action has a severe infra-red divergence,
and to perform the integral one has to regularize it
first. Let us consider the popular $\zeta$-function regularization
first. From the definition of the generalized $\zeta$-function \cite{table}
\bea
&\zeta (s,\lambda) = \dfrac{}{}\sum_{n=0}^{\infty}
\dfrac{1}{(n+\lambda)^s} 
= \dfrac{1}{\Gamma(s)} \int_0^{\infty} \dfrac{x^{s-1} \exp(-\lambda x)}
{1-\exp(-x)} dx,
\label{zeta}
\eea
we have
\bea
&\Delta {\cal L} = \dfrac{}{} \lim_{\epsilon \rightarrow 0}
\dfrac{\mu^2}{16 \pi^2} \int_{0}^{\infty} \dfrac{dt}{t^{2-\epsilon}}
\dfrac{gH}{\sinh (gHt/\mu^2)}
\Big[\exp (-2gHt/\mu^2) + \exp (+2gHt/\mu^2) \Big] \nn\\
&= \dfrac{}{} \lim_{\epsilon \rightarrow 0}
\dfrac{\mu^2}{8 \pi^2} gH \int_{0}^{\infty} \dfrac{dt}{t^{2-\epsilon}}
\dfrac{\exp (-3gHt/\mu^2) + \exp (+gHt/\mu^2)}{1-\exp(-2gHt/\mu^2)} \nn\\
&= \dfrac{}{} \lim_{\epsilon \rightarrow 0}
\dfrac{(gH)^2}{4 \pi^2} (\dfrac{2gH}{\mu^2})^{-\epsilon} \Gamma(\epsilon-1)
\Big[\zeta(\epsilon-1,\dfrac{3}{2})
+ \zeta(\epsilon-1,-\dfrac{1}{2})\Big]\nn\\
&= \dfrac{}{} \lim_{\epsilon \rightarrow 0}
\dfrac{(gH)^2}{4 \pi^2} (1-\epsilon \ln \dfrac{2gH}{\mu^2})
\big(\dfrac{1}{\epsilon} -\gamma +1 \big)
\Big[\big(\zeta(-1,\dfrac{3}{2}) + \zeta(-1,-\dfrac{1}{2})\big) \nn\\
&+ \epsilon \big(\zeta'(-1,\dfrac{3}{2})
+ \zeta'(-1,-\dfrac{1}{2})\big)\Big] \nn\\
&= \dfrac{11 g^2}{48 \pi^2} H^2 \big(\dfrac{1}{\epsilon} - \gamma
+1 - \ln \dfrac{2gH}{\mu^2} \big)
- \dfrac{g^2}{4 \pi^2} H^2 \big(2 \zeta'(-1,\dfrac{3}{2})
- i \dfrac{\pi}{2}\big),
\label{zetareg}
\eea
where $\zeta' = \dfrac{d\zeta}{ds} (s,\lambda)$,
and we have used  the fact \cite{table}
\bea
&\zeta(-1,\dfrac{3}{2}) = \zeta(-1,-\dfrac{1}{2}) = -\dfrac{11}{24},
~~~~~\zeta'(-1,-\dfrac{1}{2}) = \zeta'(-1,\dfrac{3}{2}) 
- i \dfrac{\pi}{2}. \nn
\eea
So, with the ultra-violet regularization by modified
minimal subtraction we arrive at the SNO effective
action \cite{savv,niel,ditt}
\bea
&{\cal L}_{eff}=-\dfrac{H^2}{2} -\dfrac{11g^2}{48\pi^2} H^2 (\ln
\dfrac{gH}{\mu^2}-c) + i \dfrac {g^2} {8\pi} H^2, \nn\\
&c=1-\ln 2 -\dfrac {24}{11} \zeta'(-1, \frac{3}{2})=0.94556... .
\label{snoea}
\eea
This contains the well-known imaginary part which destablizes
the SNO vacuum. Observe that
the imaginary part comes from the infra-red divergence
which originates from the tachyonic eigenstates.

\section{Monopole Background}

Notice that the SNO background (\ref{sb}) is not gauge
invariant. More seriously the separation of the SNO background
from the quantum fluctuation is not gauge independent.
So one is not sure whether the SNO effective action is
gauge independent. This is a serious defect.
To cure this defect we choose the monopole
background given by \cite{cho1,cho2}
\bea
&\vec B_\mu = \vec C_\mu,
~~~~~\vec G_{\mu\nu} = \vec H_{\mu\nu}, \nn\\
&\vec C_\mu= -\dfrac{1}{g}\hat n \times \partial_\mu\hat n, \nn\\
&\vec H_{\mu\nu}=\partial_\mu \vec C_\nu-\partial_\nu \vec C_\mu+ g
\vec C_\mu \times \vec C_\nu = -\dfrac{1}{g}
\partial_\mu\hat{n}\times\partial_\nu\hat{n},
\label{ccon}
\eea
where $\n$ is a unit isovector ($\n^2 =1$) which selects
the color charge direction everywhere in space-time.
The advantage of (\ref{ccon}) over (\ref{sb}) is that in (\ref{sb})
one can not tell the origin of the magnetic background,
whereas in (\ref{ccon}) one can tell for sure that
it comes exclusively from the non-Abelian monopole.
More importantly, here the monopole background provides
a gauge independent separation of the classical background
from the quantum fluctuation.

To see this consider the gauge-independent decomposition
of the gauge potential into the binding gluon $\hat A_\mu$
and the valence gluon $\X_\mu$ \cite{cho1,cho2},
\bea
& \vec{A}_\mu =A_\mu \n - \oneg \n\times\pro_\mu\n+\X_\mu\nonumber
         = \hat A_\mu + \X_\mu, \nn\\
& (A_\mu = \n\cdot \vec A_\mu,~~~ \hat{n}\cdot\vec{X}_\mu=0),
\label{cdec}
\eea
where $A_\mu$ is the ``electric'' potential.
Notice that $\hat A_\mu$ is precisely
the connection which leaves $\n$ invariant under parallel transport,
\bea
\D_\mu \n = \pro_\mu \n + g {\hat A}_\mu \times \n = 0.
\eea
Under the infinitesimal gauge transformation
\bea
\delta \n = - \vec \alpha \times \n  \,,\,\,\,\,
\delta \A_\mu = \oneg  D_\mu \vec \alpha,
\eea
one has
\bea
&&\delta A_\mu = \oneg \n \cdot \pro_\mu \valpha,\,\,\,\
\delta \hat A_\mu = \oneg \D_\mu \valpha  ,  \nn \\
&&\hspace{1.2cm}\delta \X_\mu = - \valpha \times \X_\mu  .
\eea
This tells that $\hat A_\mu$ by itself describes an $SU(2)$
connection which enjoys the full $SU(2)$ gauge degrees of
freedom. Furthermore $\vec X_\mu$ forms a
gauge covariant vector field under the gauge transformation.
This allows us to view QCD as the restricted gauge theory
made of the binding gluon which has the
valence gluon as the gauge covariant colored source.
But what is really remarkable is that the decomposition is
gauge independent. Once the gauge covariant topological field
$\hat n$ is chosen, the decomposition follows automatically,
regardless of the choice of gauge \cite{cho1,cho2}.

Remember that $\hat{A}_\mu$ retains all the essential
topological characteristics of the original non-Abelian potential.
First, $\hat{n}$ defines $\pi_2(S^2)$
which describes the non-Abelian monopoles.
Indeed, it is well-known that $\vec C_\mu$ with $\hn=\hat r$
describes precisely the Wu-Yang monopole \cite{wu,cho80}.
Secondly, it characterizes
the Hopf invariant $\pi_3(S^2)\simeq\pi_3(S^3)$ which describes
the topologically distinct vacua. So the topologically distinct vacua
can be described exclusively by $\hat{n}$ \cite{bpst,cho79}.
Furthermore $\hat{A}_\mu$ has a dual
structure,
\begin{eqnarray}
& \hat{F}_{\mu\nu} = (F_{\mu\nu}+ H_{\mu\nu})\hat{n}\mbox{,}
\nonumber \\
& F_{\mu\nu} = \partial_\mu A_{\nu}-\partial_{\nu}A_\mu \mbox{,}
\nonumber \\
& H_{\mu\nu} = -\dfrac{1}{g} \hat{n}\cdot(\partial_\mu
\hat{n}\times\partial_\nu\hat{n})
= \partial_\mu \tilde C_\nu-\partial_\nu \tilde C_\mu,
\end{eqnarray}
where $\tilde C_\mu$ is the ``magnetic'' potential of the monopoles
(Notice that one can always introduce the magnetic
potential since $H_{\mu \nu}$ forms a closed two-form
locally sectionwise). So the
electric-magnetic duality of QCD becomes manifest in
the restricted QCD \cite{cho1,cho2}.

Now, evidently the monopole background (\ref{ccon}) is written as
\bea
\vec H_{\mu\nu} = H_{\mu\nu}\hat n.
\eea
This demonstrates that indeed (\ref{ccon}) does describe
the gauge independent separation of the monopole
field $\vec H_{\mu\nu}$ from the generic non-Abelian
gauge field $\vec F_{\mu\nu}$. The importance of the
decomposition (\ref{cdec}) has recently been appreciated
by many authors in studying various aspects of QCD \cite{fadd,shab}.
Furthermore, in mathematics the decomposition has been shown to
play a crucial role in studying the geometry (in particular
the Deligne cohomology) of non-Abelian gauge theory \cite{cho75,zucc}.

An important feature of the decomposition (\ref{cdec})
is that it must be invariant under the color reflection \cite{cho1,cho2}
\bea
\n \rightarrow -\n,
\label{cri}
\eea
because $\hn$ is gauge equivalent to $-\hn$.
In fact $\hat A_\mu$ is explicitly invariant under the color reflection.
To understand what this means to the valence gluon $\vec X_\mu$, let
$(\hn_1, \hn_2, \hn)$ be a right-handed orthonormal basis in
$SU(2)$ space, and let
\bea
\vec X_\mu = X_\mu^1~\hn_1 +  X_\mu^2~\hn_2.
\eea
In the Abelian formalism of QCD \cite{cho3,cho4}, the valence gluon
can be expressed as a charged vector field
\bea
X_\mu = \dfrac{X_\mu^1+iX_\mu^2}{\sqrt{2}}.
\eea
Then, under the color reflection, $X_\mu$ should transform
to the charge conjugate state
\bea
X_\mu^* = \dfrac{X_\mu^1-iX_\mu^2}{\sqrt{2}}.
\eea
This amounts to changing $\vec X_\mu$ to its charge conjugate state
$\vec X_\mu^{(c)}$,
\bea
\vec X_\mu^{(c)}= X_\mu^1~\hn_1 - X_\mu^2~\hn_2,
\eea
which is equivalent to changing $\hn_2$ to $-\hn_2$.
Indeed this is exactly what we need to induce the color
reflection (\ref{cri}), because $(\hn_1, -\hn_2, -\hn)$ now
forms a right-handed orthonormal basis. This means that
the color reflection transforms  $\vec X_\mu$
to its charge conjugate state $\vec X_\mu^{(c)}$.
More importantly, $\vec X_\mu$ and $\vec X_\mu^{(c)}$
must be indistinguishable,
because they are gauge equivalent to each other.
This point will become very important in the following.

With the monopole background (\ref{ccon}) one can calculate
the functional determinant (\ref{fd}). But for the generality
we will calculate the the functional determinant with an arbitrary
(electric and magnetic) background $\hat A_\mu$.
The calculation of the ghost loop
determinant (the Faddeev-Popov determinant) $M_{ab}$ is rather
straightforward, but that of the gluon loop $K_{\mu\nu}^{ab}$
is tricky. We have
\bea
&\ln {\rm Det} K = {\rm Tr} \ln \Big(-g_{\mu\nu} \hD^2_{ab}\Big)
+ {\rm Tr} \ln \Big[g_{\mu\nu} \delta_{ab}
+ 2g G_{\mu\nu} \Big(\dfrac{N}{\hD^2}\Big)_{ab}\Big] \nn\\
&= 4 ~{\rm Tr} \ln \Big(-\hD^2_{ab}\Big)
+ {\rm Tr} \dfrac{}{}\sum_{n=1}^{\infty}
\dfrac{(-1)^{n+1}}{n} \Big(2g\dfrac{N}{\hD^2}G_{\mu\nu}\Big)^n,
\label{gdet1}
\eea
where
\bea
&G_{\mu\nu} = \partial_\mu B_\nu-\partial_\nu B_\mu
~~~(B_\mu =A_\mu + \tilde C_\mu),
~~~~~N_{ab} = \epsilon_{abc} n_c. \nn
\eea
Using the relation
\bea
&G_{\mu \alpha} G_{\nu\beta} G_{\alpha \beta} = \dfrac{1}{2} G^2 G_{\mu \nu}
+\dfrac{1}{2}(G \tilde G) {\tilde G}_{\mu \nu}
~~~~~({\tilde G}_{\mu \nu}=\dfrac{1}{2}{\epsilon}_{\mu\nu\rho\sigma}
G_{\rho\sigma}),
\eea
we can simplify the functional determinant to
\bea
& \ln {\rm Det} K = 4 ~{\rm Tr} \ln \Big(-\hD^2_{ab}\Big)
+ {\rm Tr} \ln \Big[\delta_{ab}
+4a^2\Big(\dfrac{N}{\hD^2}\Big)^2_{ab}\Big]
+{\rm Tr} \ln \Big[\delta_{ab}
-4b^2\Big(\dfrac{N}{\hD^2}\Big)^2_{ab}\Big] \nn\\
&=\ln {\rm Det} \Big[(-\hD^2+2iaN)(-\hD^2-2iaN) \Big]_{ab} \nn\\
&+\ln {\rm Det} \Big[(-\hD^2+2bN)(-\hD^2-2bN)\Big]_{ab},
\label{gdet2}
\eea
where
\bea
a = \dfrac{g}{2} \sqrt {\sqrt {G^4 + (G \tilde G)^2} + G^2},
~~~~~b = \dfrac{g}{2} \sqrt {\sqrt {G^4 + (G \tilde G)^2} - G^2}. \nn
\eea
Notice that
\bea
(-\hD^2 \pm 2iaN) \hn = 0, ~~~~~(-\hD^2 \pm 2bN) \hn = 0.
\eea
From this we can assume, without loss of generality,
that the eigenfunction of the determinants
is of the form
\bea
\vec \phi = \phi_1 \hat n_1 + \phi_2 \hat n_2.
\eea
Furthermore, we have
\bea
&\hD_\mu \hat n_1 = gB_\mu \hat n_2,
~~~~~\hD_\mu \hat n_2 = -gB_\mu \hat n_1.
\eea
With this we can simplify the eigenvalue equation
$(-\hD^2 \pm 2iaN) \vec \phi = \lambda \vec \phi$
to
\bea
\left( \begin{array}{cc}
\pro_\mu^2-g^2B_\mu^2 & -g(\pro_\mu B_\mu + 2 B_\mu \pro_\mu) \pm 2ia \\
g(\pro_\mu B_\mu + 2 B_\mu \pro_\mu) \mp 2ia & \pro_\mu^2-g^2B_\mu^2
\end{array} \right)
\left( \begin{array}{cc}
\phi_1 \\ \phi_2
\end{array} \right)
= \lambda \left(\begin{array}{cc}
\phi_1 \\ \phi_2
\end{array} \right),
\eea
which can be diagonalized to the following Abelian form
\bea
\left( \begin{array}{cc}
-\tilde D_{+}^2 \pm 2a & 0\\
0 & -\tilde D_{-}^2 \mp 2a
\end{array} \right)
\left( \begin{array}{cc}
\phi_+ \\ \phi_-
\end{array} \right)
= \lambda \left(\begin{array}{cc}
\phi_+ \\ \phi_-
\end{array} \right),
~~~~~\phi_{\pm} = \dfrac{\phi_1 \pm i\phi_2}{\sqrt{2}},
\label{evea}
\eea
where $\tilde D_{\pm}^2= (\pro_\mu \pm igB_\mu)^2$.
Similarly, we can diagonalize the equation
$(-\hD^2 \pm 2bN) \vec \phi = \lambda \vec \phi$ to
\bea
\left( \begin{array}{cc}
-\tilde D_{+}^2 \mp 2ib & 0 \\
0 &-\tilde D_{-}^2 \pm 2ib
\end{array} \right)
\left( \begin{array}{cc}
\phi_+ \\ \phi_-
\end{array} \right)
= \lambda \left(\begin{array}{cc}
\phi_+ \\ \phi_-\end{array} \right).
\label{eveb}
\eea
From this we have
\bea
&{\rm Det} (-\hD^2+2iaN)_{ab}
= {\rm Det} (-\tilde D_{+}^2+2a)(-\tilde D_{-}^2-2a), \nn\\
&{\rm Det} (-\hD^2-2iaN)_{ab}
= {\rm Det} (-\tilde D_{+}^2-2a)(-\tilde D_{-}^2+2a), \nn\\
&{\rm Det} (-\hD^2+2bN)_{ab}
= {\rm Det} (-\tilde D_{+}^2-2ib)(-\tilde D_{-}^2+2ib), \nn\\
&{\rm Det} (-\hD^2-2bN)_{ab}
= {\rm Det} (-\tilde D_{+}^2+2ib)(-\tilde D_{-}^2-2ib).
\label{agdet}
\eea
At this point it is important to realize that
$(-\tilde D_{-}^2 \mp 2a)$ and
$(-\tilde D_{-}^2 \pm 2ib)$ are what one obtains from
$(-\tilde D_{+}^2 \pm 2a)$ and
$(-\tilde D_{+}^2 \mp 2ib)$ by replacing $g$ to $-g$,
so that they are charge conjugate to the others.
Furthermore, the two eigenfunctions
$\phi_{+}$ and $\phi_{-}$ are also
charge conjugate to each other (although they are not
complex conjugate to each other). To see this notice that
they are related by changing $\phi_2$ to $-\phi_2$.
This, viewed in terms of $\vec \phi$,
amounts to changing $\hn_2$ to $-\hn_2$. But,
as we have noted before, this change is equivalent
to the change of the color direction $\hn$ to $-\hn$.
This is nothing but the color reflection (\ref{cri}).
This means that $(-\tilde D_{+}^2 \pm 2a)\phi_+$ and
$(-\tilde D_{-}^2 \mp 2a)\phi_-$, and $(-\tilde D_{+}^2 \mp 2ib)\phi_+$
and $(-\tilde D_{-}^2 \pm 2ib)\phi_-$, are
not only charge conjugate to each other, but also gauge
equivalent to each other.
This means that they must have
identical eigenvalues, so that
\bea
&{\rm Det} (-\tilde D_{+}^2\pm2a) = {\rm Det} (-\tilde D_{-}^2\mp2a), \nn\\
&{\rm Det} (-\tilde D_{+}^2\mp2ib) = {\rm Det} (-\tilde D_{-}^2\pm2ib).
\label{ccagdet}
\eea
From this we have
\bea
&\ln {\rm Det} (-\hD^2\pm2iaN)_{ab}
= 2 \ln {\rm Det} (-\tilde D_{+}^2\pm2a), \nn\\
&\ln {\rm Det} (-\hD^2\pm2bN)_{ab} = 2 \ln {\rm Det} (-\tilde D_{+}^2\mp2ib).
\label{cci}
\eea
This tells that we can reduce
the eigenvalue problem of a non-Abelian
functional determinants $K_{\mu\nu}^{ab}$ and $M_{ab}$
to the eigenvalue problem of the following Abelian determinants,
\bea
&\ln {\rm Det} K = 2 \ln {\rm Det} (-\tilde D^2+2a)(-\tilde D^2-2a) \nn\\
&+2 \ln {\rm Det} (-\tilde D^2-2ib)(-\tilde D^2+2ib), \nn\\
&\ln {\rm Det} M = 2 \ln {\rm Det} (-\tilde D^2),
\label{agdetx}
\eea
where
\bea
\tilde D_\mu = \pro_\mu +igB_\mu. \nn
\eea
Notice that in the Lorentz frame where the
electric field becomes parallel to the magnetic field, $a$ becomes
purely magnetic and $b$ becomes purely electric.

From this we obtain \cite{ditt,cho3,cho4}
\bea
&\Delta S = i \ln {\rm Det} (-\tilde D^2+2a)(-\tilde D^2-2a)
+ i \ln {\rm Det} (-\tilde D^2-2ib)(-\tilde D^2+2ib) \nn\\
&- 2i \ln {\rm Det}(-\tilde D^2),
\label{fdabx}
\eea
and
\bea
&\Delta {\cal L} =  \dfrac{}{} \lim_{\epsilon\rightarrow0}
\dfrac{1}{16 \pi^2}  \int_{0}^{\infty}
\dfrac{dt}{t^{3-\epsilon}} \dfrac{ab t^2}{\sinh (at/\mu^2)
\sin (bt/\mu^2)} \Big[ \exp(-2at/\mu^2)+\exp(+2at/\mu^2) \nn\\
&+\exp(+2ibt/\mu^2)+\exp(-2ibt/\mu^2)-2 \Big].
\label{eaabx}
\eea
Notice that for the monopole background (\ref{ccon}) we have $b=0$,
so that the integral (\ref{eaabx}) is reduced to
\bea
&\Delta{\cal L} = \dfrac{}{} \lim_{\epsilon\rightarrow0}
\dfrac{1}{16 \pi^2}\int_{0}^{\infty}
\dfrac{dt}{t^{2-\epsilon}} \dfrac{a \mu^2}{\sinh (at/\mu^2)}
\Big[\exp (-2at/\mu^2 ) + \exp (+2at/\mu^2) \Big].
\label{eaax}
\eea
With $a=gH$ this reduces to (\ref{eahx}).

\section{Infra-red Regularization by Causality}

The integral (\ref{eaabx}) has a severe infra-red
divergence. This is due to the fact that ${\rm Det}(-\tilde D^2-2a)$
and ${\rm Det}(-\tilde D^2+2ib)$ have unstable eigenstates.
To see this notice that, to calculate the eigenvalues,
one often has to choose a particular gauge and a particular
Lorentz frame. A best way to calculate
the determinants is to go to the gauge and Lorentz frame
where the color electromagnetic field assumes
a particular direction. In this gauge one can easily show that
${\rm Det}(-\tilde D^2-2a)$ has negative eigenvalues
and thus tachyonic eigenstates when $k^2<2a$, where $k$ is the momentum
of the eigenstate in the direction of the color magnetic field \cite{niel}.
Similarly, ${\rm Det}(-\tilde D^2+2ib)$ has
acausal eigenstates which propagate backward in time when $k^2<2b$.
These acausal states create the instability which causes
the infra-red divergence in (\ref{eaabx}) \cite{ditt}.

Notice, however, that
there is a subtle but very important point that we have
overlooked in the calculation of the determinants (\ref{fdhx})
and (\ref{fdabx}). The classical backgrounds (\ref{sb}) and (\ref{ccon})
must be gauge invariant, so that in the calculation of
the functional determinant (\ref{gdet2}) we should make
sure that the gauge invariance, in particular
the color reflection invariance, is properly implemented.
Unfortunately we did not. When we do this correctly, the acausal tachyonic
states disappear from the physical eigenstates of the determinants,
and the effective action no longer has the infra-red divergence.
Before we discuss how this happens in Section VII in detail
we show that we still have a chance to correct this mistake
when we make the infra-red regularization.

The evaluation of the integral (\ref{eaabx}) for arbitrary
$a$ and $b$ has been notoriously difficult \cite{savv,niel,ditt}.
Even in the case of ``simpler'' QED, the integration of the
effective action has been completed only recently \cite{cho01,cho5}.
Fortunately the integral for the pure electric ($a=0$) and
pure magnetic ($b=0$) background can
be correctly performed \cite{cho3,cho4}.

Consider the pure magnetic background (\ref{eaax}) first.
To perform the integral we have to regularize it
first. There are two competing infra-red regularizations,
the standard $\zeta$-function regularization and the regularization by
causality. Since we have already discussed the $\zeta$-function
regularization, here we review
the infra-red regularization by causality.
For this we go to the Minkowski time with
the Wick rotation, and find \cite{cho3,cho4}
\bea
\label{eaam}
&\Delta {\cal L}=  \Delta{\cal L_+} + \Delta{\cal L_-}, \nn\\
& \Delta{\cal L_+} =  - \dfrac{}{} \lim_{\epsilon \rightarrow 0}
\dfrac{1}{16 \pi^2}\int_{0}^{\infty}
\dfrac{dt}{t^{2-\epsilon}} \dfrac{a \mu^2}{\sin (at/\mu^2)}
\exp (-2i a t/\mu^2 ), \nn\\
& \Delta{\cal L_-} =  - \dfrac{}{} \lim_{\epsilon \rightarrow 0}
\dfrac{1}{16 \pi^2}\int_{0}^{\infty}
\dfrac{dt}{t^{2-\epsilon}} \dfrac{a\mu^2}{\sin (at/\mu^2)}
\exp (+2i a t/\mu^2 ).
\eea
In this form the infra-red divergence has disappeared,
but now we face an ambiguity in choosing the correct contours
of the integrals in (\ref{eaam}). Fortunately this ambiguity can
be resolved by causality. To see this notice that the two integrals
$\Delta{\cal L_+}$ and $\Delta{\cal L_-}$ originate from the
two determinants $(-\tilde D^2+2a)$ and $(-\tilde D^2-2a)$,
and the standard causality argument (with the familiar
Feynman prescription $p^2 \rightarrow p^2-i\epsilon$)
requires us to identify $2 a$ in the first determinant as
$2 a -i\epsilon$ but in the second determinant as
$2 a +i\epsilon$. This tells that
the poles in the first integral in (\ref{eaam}) should lie above
the real axis, but the poles in the second integral should lie
below the real axis. From this we conclude
that the contour in $\Delta{\cal L_+}$ should pass below the
real axis, but the contour in $\Delta{\cal L_-}$ should pass above the
real axis. With this causality requirement the two integrals
become complex conjugate to each other. This guarantees that
$\Delta{\cal L}$ is explicitly real, without any imaginary part.
This suggests that in the Euclidian time $\Delta{\cal L_+}$
and $\Delta{\cal L_-}$ must have exactly the same expression,
\bea
&\Delta{\cal L_+} = \Delta{\cal L_-} 
= \dfrac{}{} \lim_{\epsilon \rightarrow 0}
\dfrac{1}{16 \pi^2}\int_{0}^{\infty}
\dfrac{dt}{t^{2-\epsilon}} \dfrac{a\mu^2}{\sinh (at/\mu^2)} 
\exp (-2a t/\mu^2).
\eea
We will see that this is exactly what one obtains
when one calculates the functional determinant (\ref{gdet1})
correctly.

With this infra-red regularization by causality we
obtain \cite{cho3}
\bea
&{\cal L}_{eff} = - \dfrac{a^2}{2g^2} -\dfrac{11a^2}{48\pi^2}(\ln
\dfrac{a}{\mu^2}-c),
\label{ceaa}
\eea
for a pure monopole background.
This is identical to the SNO effective action (\ref{snoea}),
except that it no longer contains the imaginary part.

For the pure electric background (i.e., with $a=0$)
(\ref{eaabx}) is reduced to
\bea
&\Delta {\cal L}  =  \dfrac{}{} \lim_{\epsilon \rightarrow 0}
\dfrac{1}{16 \pi^2}  \int_{0}^{\infty}
\dfrac{ d t}{t^{2-\epsilon}} \dfrac{b\mu^2}{\sin (bt/\mu^2)} 
\Big[\exp(+2ibt/\mu^2) +\exp(-2ibt/\mu^2) \Big],
\label{eabx}
\eea
so that, with the infra-red regularization by causality,
we obtain \cite{cho3,cho4}
\bea
\label{ceab}
&{\cal L}_{eff} = \dfrac{b^2}{2g^2} +\dfrac{11b^2}{48\pi^2}
(\ln \dfrac{b}{\mu^2}-c) -i\dfrac{11b^2}{96\pi}.
\eea
From this we obtain
\bea Im ~\Delta {\cal L} = \left\{{~~~~0~~~~~~~~~~~~(b=0),
\atop -\dfrac{11 b^2}{96 \pi} ~~~~~~~~~(a=0).}\right.
\label{imea}
\eea
Notice that when $a=0$, the imaginary part has a negative
signature.

Obviously the difference between (\ref{snoea}) and (\ref{ceaa})
originates from the different infra-red regularizations.
The question now is which regularization,
the $\zeta$-function regularization or the infra-red regularization
by causality, is the correct one.

\section{Perturbative Calculation of the Imaginary Part}

Fortunately we can answer this question with a perturbative
method. This is because in massless gauge theories (in particular
in QCD) the imaginary part of the effective
action is of the order of $g^2$ \cite{sch,cho4}.
We first demonstrate that one can indeed calculate
the imaginary part of the effective
action perturbatively in massless QED,
and apply the perturbative method to obtain the imaginary part
of the QCD effective action.

To do this we review the Schwinger's
perturbative calculation of the QED effective action. In QED
Schwinger has obtained the following effective action
perturbatively to the order $e^2$ \cite{schw}
\bea
&\Delta S_{QED}=\dfrac{e^2}{16 \pi^2} \int d^4p
F_{\mu\nu}(p)F_{\mu\nu}(-p)
\dfrac{}{}\int_{0}^{1} dv \dfrac{v^2 (1- v^2/3)}{(1- v^2)
+ 4m^2/p^2},
\label{qedea}
\eea
where $m$ is the electron mass. From this he observed that
when $-p^2>4m^2$ the integrand develops a pole at
$v^2=1+4m^2/p^2$ which generates an imaginary part, and
explained how to calculate the imaginary part
of the effective action. But notice that in the massless limit,
the pole moves to $v=1$. In this case the pole contribution
to the imaginary part is reduced by a half, and we obtain
\bea
Im ~{\cal L}_{QED} {\Big |}_{m=0} = \left\{{~0
~~~~~~~~~~~~~~ b=0,
\atop \dfrac{b^2}{48 \pi} ~~~~~~~~~~~a=0.}\right.
\eea
This is exactly what we obtain
from the non-perturbative effective action in
the massless limit \cite{cho01,cho5}.
This confirms that in massless QED, one can calculate the
imaginary part of the effective action either perturbatively or
non-perturbatively, with identical results.

\begin{figure*}
\begin{center}
\includegraphics{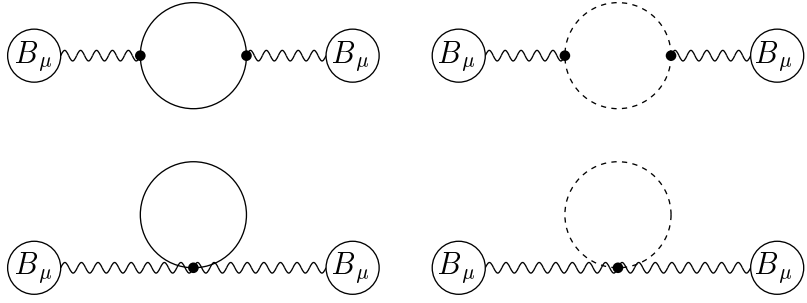}
\end{center}
\caption{\small{\label{Fig. 2} The Feynman diagrams that
contribute to the effective action at $g^2$ order. Here the
straight line and the dotted line represent the valence gluon and
the ghost, respectively.}}
\end{figure*}

Now we repeat the perturbative calculation for QCD.
We can do this either by calculating
the one-loop Feynman diagrams directly,
or by evaluating the integral (\ref{ea}) perturbatively to
the order $g^2$. We start with the Feynman diagrams.
For an arbitrary background $B_\mu$ there are four
Feynman diagrams that contribute to the order $g^2$ which are shown
in Fig. 2. Notice that the tadpole diagrams contain a quadratic
divergence which does not appear in the final result.

The sum of these diagrams (in the Feynman gauge with dimensional
regularization) gives us \cite{pesk}
\bea \label{peafeyn}
&\Delta S = -\dfrac{11g^2}{96 \pi^2} \int d^4p
G_{\mu\nu}(p)G_{\mu\nu}(-p) 
\left[\mbox{ln}
\left(\dfrac{p^2}{\mu^2}\right) + C_1 \right],
\eea
where $C_1$ is a regularization-dependent constant.
Clearly the imaginary part could only
arise from the logarithmic term $\mbox{ln} (p^2/\mu^2)$, so that
for a space-like $p^2$ (with $\mu^2>0$) the effective action
has no imaginary part. However, since a space-like
$p^2$ corresponds to a magnetic background, we find that
the magnetic condensation generates no imaginary part,
at least at the order $g^2$. To evaluate the imaginary part
for an electric background we have to make the analytic
continuation of (\ref{peafeyn}) to a time-like $p^2$,
because the electric background corresponds to a time-like $p^2$.
In this case the causality (again with the Feynman
prescription $p^2 \rightarrow
p^2-i \epsilon$) dictates us to have
\bea
&\mbox{ln} \Big(\dfrac{p^2}{\mu^2}\Big) \rightarrow
\lim_{\epsilon \to0} \mbox{ln} \Big(\dfrac{p^2-i\epsilon}{\mu^2} \Big) 
=\mbox{ln} \Big(\dfrac{|p^2|}{\mu^2}\Big)- i \dfrac{\pi}{2}
~~~~~(p^2<0),
\label{causal}
\eea
so that we obtain
\bea Im ~\Delta {\cal L} = \left\{{~~~~0~~~~~~~~~~~~(b=0),
\atop -\dfrac{11 b^2}{96 \pi} ~~~~~~~~~(a=0).}\right. \nn
\eea
Obviously this is identical to (\ref{imea}). This allows us
to conclude that the result (\ref{imea}) is
indeed endorsed by the Feynman diagram calculation.

To remove any lingering doubt about (\ref{imea}) we now make the
perturbative calculation of the integral (\ref{ea}) to the order
$g^2$ with the Schwinger's method, and find \cite{cho4,hon}
\bea
&\Delta S = -\dfrac{g^2}{8\pi^2}
\int d^4p G_{\mu \nu}(p) G_{\mu \nu}(-p) \Sigma (p), \nn\\
&\Sigma (p) = \dfrac{}{}\int_{0}^{1} dv (1-\dfrac{v^2}{4})
\dfrac{}{}\int_{0}^{\infty}\dfrac{dt}{t}
\exp[-\dfrac{p^2}{4}(1-v^2) t] \nn\\
&= 2\dfrac{}{}\int_{0}^{1} dv \dfrac{v^2(1-v^2/12)}{1-v^2} + C_2,
\label{peasch}
\eea
where $C_2$ is a regularization-dependent constants.
Now, it is straightforward to evaluate the imaginary part of $\Delta S$.
Comparing this with Schwinger's result (\ref{qedea}) for
the massless QED we again reproduce (\ref{imea}),
after the proper charge and wave function
renormalization.

Furthermore, from the definition of the exponential
integral function \cite{table}
\bea
&Ei (-z) = - \dfrac{}{} \int_{z}^{\infty} \dfrac{d\tau}{\tau}
\exp (-\tau) = \gamma + \mbox{ln} z \nn\\
&+ \dfrac{}{} \int_{0}^{z} \dfrac{d\tau}{\tau}
\left[\exp (-\tau)-1\right]~~~({\rm Re}~z > 0),
\eea
we can express $\Sigma (p)$ as
\bea
&\Sigma (p) = - \dfrac{}{}\lim_{\epsilon\to0} \int_{0}^{1} dv
(1-\dfrac{v^2}{4}) \left[ \gamma + \mbox{ln} \Big(\dfrac{p^2}{4 \mu^2}(1-v^2)
\epsilon \Big) \right] \nn\\
&= \dfrac{11}{12}\left[\mbox{ln}\left(\dfrac{p^2}{\mu^2}\right) + C_3 \right],
\label{sigma}
\eea
where $C_3$ is another
regularization-dependent constant.
This tells that (\ref{peasch}) is identical to (\ref{peafeyn}).
This is the reason why the perturbative calculation by Feynman diagrams
and by Schwinger's method produce the same result.
This strongly indicates that the monopole condensation
indeed describes a stable vacuum, but the electric background creates
the pair-annihilation of the valence gluons in
$SU(2)$ QCD \cite{cho3,cho4}.

It might look surprising that both
the infra-red regularization by causality and
the perturbative method endorse the stability of the monopole
condensation. But this is not accidental. In fact this is
natural because both are based on
the causality, which is what we need
to exclude the unphysical tachyonic modes.

This confirms that the tachyonic modes are indeed unphysical
mirage which should not have been there in the first place.
They come into the calculation of the functional determinant
by default. If so, one may ask how one can prevent this.
Now we prove that a proper implementation of the gauge invariance
excludes the tachyonic modes
from the functional determinant.

\section{Color Reflection Invariance:
A Gauge Invariant Calculation of Functional Determinant}

The effective action (\ref{ea}) is nothing but the vacuum to vacuum
amplitude in the presence of the external field $\vec B_\mu$ \cite{pesk},
\bea
&\exp \Big[i S_{eff} (\vec B_\mu)\Big]
= <\Omega_+|~\Omega_-> \Big|_{\vec B_\mu} \nn\\
&= \dfrac{}{} \sum_{|n_i>}<\Omega_+|~n_i><n_i~|~\Omega_-> \Big|_{\vec B_\mu},
\label{vtv}
\eea
where $|\Omega>$ is the vacuum and $|n_i>$ is
a complete set of orthonormal states
of QCD. In the integral expression (\ref{ea}) the summation
in terms of the complete set of states is expressed
by the functional integration (at one-loop level).
Now, to calculate this vacuum to vacuum amplitude
one must use the physical vacuum and physical spectrum.
And clearly the physical spectrum should not include
the unphysical tachyons, unless one wants to assert that
the physical spectrum of QCD must contain the tachyonic modes.
This means that $|n_i>$ should not include
the unphysical acausal states.
This justifies the exclusion of the unphysical states
in the calculation of the effective action.
The question now is how one can do so.

There are two ways to exclude
the unphysical modes, when
one calculates the functional determinant
or when one makes the infra-red regularization.
We have already shown how to do this when we make
the infra-red regularization \cite{cho1,cho2}.

Now we are ready to show that a correct evaluation of
the functional determinant (\ref{gdet2}) automatically excludes
the tachyonic modes. In fact we will show that
the gauge invariance forbids them to qualify as the physical
eigenstates of the functional determinant (\ref{gdet2}).
In the absence of the tachyonic modes, of course,
we have no infra-red divergence,
and thus no need of any infra-red regularization.
This tells that it is not the $\zeta$-function
regularization, but the incorrect evaluation of the functional
determinant (\ref{gdet2}), which causes the instability of
the SNO vacuum.

When one evaluates the functional determinant (\ref{gdet2}),
one must start from
a gauge invariant background. Otherwise the effective action
will not remain gauge invariant. This means that the backgrounds
(\ref{sb}) and (\ref{ccon}) must be invariant
under the background gauge transformation (\ref{bgt}).
But obviously the background (\ref{sb}) is not
gauge invariant, so that one must impose the gauge invariance
by hand in the calculation of the functional determinant (\ref{fdhx}).
Of course, one could choose a particular gauge to calculate
the determinant, but one should make sure that the gauge invariance
is properly implemented in the calculation.

To discuss the gauge invariance it is important to understand
the color reflection invariance in QCD \cite{cho1,cho2}.
Consider the decomposition (\ref{cdec}) again.
As we have already noted, the decomposition (\ref{cdec})
can be defined only up to the color reflection (\ref{cri}).
This tells that the electric potential $A_\mu$ in (\ref{cdec}),
and thus the color charge in QCD, can uniquely be defined only up to
this reflection degrees of freedom, even after one has chosen one's
color direction \cite{cho1,cho2}. Furthermore, this means that
any physical state must be invariant
under the the color reflection group generated by (\ref{cri}).
To understand this in physical terms, consider a $q \bar q$ state.
There are two color neutral states
\bea
&|C,C_3> = |0,0> = \dfrac {|r\bar r>+~|b\bar b>}{\sqrt 2}, \nn\\
&|C,C_3> = |1,0> = \dfrac {|r\bar b>-~|b\bar r>}{\sqrt 2},
\eea
where $r$ and $b$ represent the red and blue quark.
But obviously only the first one is color singlet, and thus
qualifies as a physical state.
Now, notice that under the color reflection (\ref{cri})
the role of red and blue colors are interchanged,
and only the color singlet state remains invariant
under the color reflection. This shows that the
color reflection invariance is an important
symmetry, a prerequisite for a physical state, of QCD \cite{cho1,cho2}.
In $SU(2)$ (\ref{cri}) generates a 4-element color reflection group,
and in $SU(3)$ the color reflection group forms a 24-element
symmetry group.

Now let us go back to the Savvidy background (\ref{sb}) again
and show that the gauge invariance
excludes the tachyonic modes from the physical
eigenstates. To implement the gauge invariance to the Savvidy background,
notice that $\vec G_{\mu\nu}$ in (\ref{sb})
must be gauge covariant. So one should be able
to change $\vec G_{\mu\nu}$ to
$-\vec G_{\mu\nu}$ by a gauge transformation by rotating
$\hn_0$ to $-\hn_0$. In terms of $\bar H_{\mu\nu}$,
this means that one can change $\bar H_{\mu\nu}$
to $-\bar H_{\mu\nu}$ by a gauge transformation.
But obviously the gluon spin is not affected by this gauge
transformation. {\it This means that one can change the spin
polarization direction of the gluon with respect to
the magnetic field by a gauge transformation, which tells that
the gauge invariant magnetic background can not have
a polarization direction}.
Consequently the gauge invariant eigenvalues of
the equation (\ref{eveh}) must be those which are invariant under
the transformation $H$ to $-H$. This, in turn, means that
only the eigenstates which are invariant under the spin flip
$(S_3 \rightarrow -S_3)$ of the valence gluon
can be treated as the physical states.
This is nothing but
the color reflection invariance. Now, it must be clear that
the tachyonic states (more precisely the eigenstates with $S_3=-1$
and $n=0$) are precisely the states which violate this color
reflection invariance. This is shown schematically in Fig. 1,
where (A) transforms to (B) under the color
reflection. Notice that only the eigenstates with positive
eigenvalues remain invariant under the color reflection.
This tells that one must exclude
the tachyonic states in one's calculation of
the effective action (\ref{fdhx}).

If one does so,
the effective action (\ref{fdhx}) changes to
\bea
&\Delta S = i\ln {\rm Det} \Big[(-\bD^2+2gH)(-\bD^2+2gH) \Big],
\label{fdho}
\eea
and we have
\bea
&\Delta{\cal L} = \dfrac{}{} \lim_{\epsilon \rightarrow 0}
\dfrac{1}{16 \pi^2}\int_{0}^{\infty}
\dfrac{dt}{t^{2-\epsilon}} \dfrac{gH/ \mu^2}{\sinh (gHt/\mu^2)} 
\Big[\exp (-2gHt/\mu^2 )
+ \exp (-2gHt/\mu^2) \Big],
\label{eaho}
\eea
which has no infra-red divergence.
This precludes the necessity to make any infra-red regularization.

The fact that $\bar H_{\mu\nu}$ does not change
to $-\bar H_{\mu\nu}$ under the reflection
$\n_0$ to $-\n_0$ in (\ref{sb}) simply confirms that the Savvidy background
is not gauge invariant. This must be compared with
the monopole background (\ref{ccon}), which clearly
has the advantage that it is invariant under
the color reflection (\ref{cri}).
Indeed not only $\vec C_\mu$ but also the background
$\hat A_\mu$ itself is invariant
under the color reflection (because the electric
potential $A_\mu$ is uniquely defined only up to the signature).

Let us calculate the functional determinant (\ref{gdet2})
with this monopole background (\ref{ccon}).
Clearly we must honor the gauge invariance in the calculation.
Unfortunately we did not. To see this consider
${\rm Det} (-\hD^2+2iaN)(-\hD^2-2iaN)$
in (\ref{gdet2}) first. This is made of two determinants,
\bea
{\rm Det} (-\hD^2+2iaN)={\rm Det} (-\tilde D^2+2a)^2, \nn\\
{\rm Det} (-\hD^2-2iaN)={\rm Det} (-\tilde D^2-2a)^2,
\eea
and we have to solve the eigenvalue problem for
each determinant separately. At first sight it appears that
only the second determinant contains the tachyonic eigenstates.
However, observe that $\hD^2$ is invariant under (\ref{cri}),
but $N_{ab}$ changes the signature. So the two determinants are
the color reflection counterpart of each other. This means that,
after the color reflection, it becomes the first determinant
which contains the tachyonic eigenstates when $k^2<a$.
From this it must become clear that actually both determinants
contain the tachyonic eigenstates. {\it More importantly,
in both determinants, the tachyonic eigenstates are precisely those
which do not remain invariant
under the color reflection. Furthermore, the color
reflection invariance tells that the two determinants
must have identical eigenvalues. This means
that the physical eigenstates of
${\rm Det} (-\hD^2+2iaN)(-\hD^2-2iaN)$ must be given by
\bea
{\rm Det} (-\tilde D^2+2a)^2(-\tilde D^2+2a)^2,
\eea
not by
\bea
{\rm Det} (-\tilde D^2+2a)^2(-\tilde D^2-2a)^2,
\eea
and should not
contain any tachyonic states}.

Exactly the same argument applies to
${\rm Det} (-\hD^2+2bN)(-\hD^2-2bN)$ in (\ref{gdet2}),
made of two determinants
\bea
{\rm Det} (-\hD^2+2bN)={\rm Det} (-\tilde D^2-2ib)^2, \nn\\
{\rm Det} (-\hD^2-2bN)={\rm Det} (-\tilde D^2+2ib)^2.
\eea
In this case only the second determinant appears to have acausal
eigenstates propagating backward in time when $k^2<b$,
where $k$ is the momentum of eigenstate in the direction of
the background color electric field. But again, the above argument
tells that the two determinants are the color reflection
counterpart of each other. So, under the color reflection
the role of the first and second determinants are interchanged,
and only the causal eigenstates which become invariant
under the color reflection qualify as the physical eigenstates.
From this we conclude that ${\rm Det} (-\hD^2+2bN)(-\hD^2-2bN)$
should be written as
\bea
{\rm Det} (-\tilde D^2-2ib)^2(-\tilde D^2-2ib)^2,
\eea
not as
\bea
{\rm Det} (-\tilde D^2-2ib)^2(-\tilde D^2+2ib)^2.
\eea
So, excluding the unphysical eigenstates we have
\bea
& \ln {\rm Det} K 
= \ln {\rm Det} \Big[(-\hD^2+2iaN)(-\hD^2+2iaN) \Big]_{ab} \nn\\
&+\ln {\rm Det} \Big[(-\hD^2+2bN)(-\hD^2+2bN)\Big]_{ab} \nn\\
&= 2 \ln {\rm Det} (-\tilde D^2+2a)(-\tilde D^2+2a)
+2 \ln {\rm Det} (-\tilde D^2-2ib)(-\tilde D^2-2ib).
\label{gdeto}
\eea
From this we finally have
\bea
&\Delta S = i \ln {\rm Det} [(-\tilde D^2+2a)(-\tilde D^2+2a)]
+ i \ln {\rm Det} [(-\tilde D^2-2ib)(-\tilde D^2-2ib)] \nn\\
&- 2i \ln {\rm Det}(-\tilde D^2),
\label{fdabo}
\eea
and
\bea
&\Delta {\cal L} =  \dfrac{}{} \lim_{\epsilon\rightarrow0}
\dfrac{1}{16 \pi^2}  \int_{0}^{\infty}
\dfrac{dt}{t^{3-\epsilon}} \dfrac{ a b t^2 / \mu^4}{\sinh (at/\mu^2)
\sin (bt/\mu^2)} \Big[ \exp(-2at/\mu^2)+\exp(-2at/\mu^2) \nn\\
&+\exp(+2ibt/\mu^2)+\exp(+2ibt/\mu^2)-2 \Big].
\label{eaabo}
\eea
Obviously this has no infra-red divergence.
This tells that, only when one mistakenly includes the
unphysical eigenstates in the calculation of the functional
determinant, one has to worry about which infra-red
regularization is the correct one.

In retrospect we could have asserted that the tachyonic modes
should be excluded from the functional determinant
simply by saying that they are unphysical,
because violate the causality. Clearly this is a correct assertion.
But without an independent justification of this
one might have objected this assertion. The above analysis
tells that we can actually make this assertion, because
we now have an independent verification
of this assertion with the gauge invariance.

At this point we must clarify the meaning of the gauge invariant
background. By this
we do not mean that the background $\vec G_{\mu\nu}$ is
gauge invariant. It must be gauge covariant. By a gauge invariant
background we mean $\vec G_{\mu\nu}^2$ of a gauge covariant
$\vec G_{\mu\nu}$. Exactly the same interpretation applies to
the Lorentz invariance. In this sense our vacuum made of
the monopole condensation is clearly gauge invariant
as well as Lorentz invariant. Of course one could choose
any gauge one likes to calculate the functional determinant
(\ref{gdet2}). For example for the monopole background
one can certainly choose a gauge where $\n$ becomes
$\n_0$. When one does that the monopole background
becomes $H_{\mu\nu} \n_0$, which looks almost identical to
the Savvidy background $\bar H_{\mu\nu} \n_0$. But we emphasize that
even in this gauge it remains invariant under the color reflection,
because $H_{\mu\nu}$ transforms to $-H_{\mu\nu}$
under (\ref{cri}).

Now, one can integrate (\ref{eaabo}) and obtain \cite{cho3,cho4}
\bea
{\cal L}_{eff}=\left\{\begin{array}{ll}-\dfrac{a^2}{2g^2}
-\dfrac{11a^2}{48\pi^2}
(\ln \dfrac{a}{\mu^2}-c),~~~~~b=0 \\
~\dfrac{b^2}{2g^2} +\dfrac{11b^2}{48\pi^2}
(\ln \dfrac{b}{\mu^2}-c) \\
-i\dfrac{11b^2}{96\pi},
~~~~~~~~~~~~~~~~~~~~~~~~~~~~a=0  \end{array}\right.
\label{ceaab}
\eea
which is identical to the effective action that we obtained
with the infra-red regularization by causality.
It is really remarkable that two completely independent
principles, the gauge invariance and the causality,
produce exactly the same effective action in QCD.

Observe that the effective action (\ref{ceaab}) with $a=0$
and $b=0$ are related by the electric-magnetic
duality \cite{cho3,cho01}.
In fact we can obtain one from the other
simply by replacing $a$ with $-ib$ and $b$ with $ia$.
This duality, which states that the effective action should be
invariant under the replacement
\bea
a \rightarrow - ib,~~~~~~~b \rightarrow ia,
\label{dt}
\eea
was first discovered in the effective action of
QED recently \cite{cho01,cho5}.
But subsequently this duality has also been shown to exist
in the QCD effective action \cite{cho3}. This tells that
the electric-magnetic duality should be regarded as a
fundamental symmetry of the effective action of gauge theory,
both Abelian and non-Abelian. The importance of this duality
is that it provides
a very useful tool to check the self-consistency
of the effective action. The fact that (\ref{ceaa})
and (\ref{ceab}) are related by the duality
assures that the infra-red regularization by causality
is self-consistent.

The effective action (\ref{ceaab}) generates the much desired
dimensional transmutation in QCD, the phenomenon
Coleman and Weinberg first observed in massless scalar QED \cite{cole}.
To demonstrate this we first renormalize the effective
action. Notice that the effective action (\ref{ceaab})
provides the following effective potential
\bea
V=\dfrac{a^2}{2g^2}
\Big[1+\dfrac{11 g^2}{24 \pi^2}(\ln\dfrac{a}{\mu^2}-c)\Big].
\eea
So we define the running coupling $\bar g$ by \cite{savv,cho3}
\bea
\frac{\partial^2V}{\partial a^2}\Big|_{a=\bar \mu^2}
=\frac{1}{ \bar g^2}.
\eea
With the definition we find
\bea
\frac{1}{\bar g^2} =
\frac{1}{g^2}+\frac{11}{24 \pi^2}( \ln\frac{{\bar\mu}^2}{\mu^2}
- c + \dfrac{3}{2}),
\eea
from which we obtain the following $\beta$-function,
\bea
\beta(\bar\mu)= \bar\mu \dfrac{\partial \bar g}{\partial \bar\mu}
= -\frac{11 \bar g^3}{24\pi^2}.
\eea
This is exactly the same $\beta$-function that one obtained
from the perturbative QCD to prove the asymptotic freedom
\cite{wil}. The fact that the $\beta$-function obtained from
the effective action becomes identical to the one obtained by
the perturbative calculation is remarkable, because
this is not always the case. In fact in QED it has been
shown that the running coupling and the $\beta$-function
obtained from the effective action is different from those
obtained from the perturbative method \cite{cho01,cho5}.

\begin{figure*}\begin{center}
\includegraphics{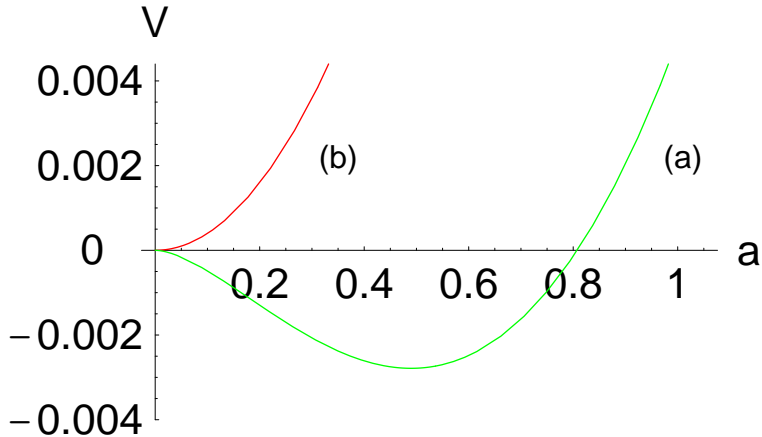}\end{center}
\caption{\label{Fig. 3} The effective potential of SU(2) QCD in
the pure magnetic background. Here (a) is the effective potential
and (b) is the classical potential.}
\end{figure*}

In terms of the running coupling the renormalized potential is given by
\bea
V_{\rm ren}=\dfrac{a^2}{2\bar g^2}
\Big[1+\dfrac{11 \bar g^2}{24 \pi^2 }
(\ln\dfrac{a}{\bar\mu^2}-\dfrac{3}{2})\Big],
\eea
which generates a non-trivial local minimum at
\bea
<a>=\bar \mu^2 \exp\Big(-\frac{24\pi^2}{11\bar g^2}+ 1\Big).
\eea
Notice that with ${\bar \alpha}_s = 1$ we have
\bea
\dfrac{<a>}{{\bar \mu}^2} = 0.48988... .
\eea
This is nothing but the desired magnetic condensation.
The corresponding effective potential is plotted in Fig. 3,
where we have assumed $\bar \alpha_s = 1$ and $~\bar \mu =1$.

Nelsen and Olesen have suggested that the existence
of the unstable tachyonic modes are closely related with
the asymptotic freedom in QCD \cite{niel}.  Our analysis tells
that this is not true. Obviously our asymptotic freedom
follows from a stable monopole condensation.

\section{Discussion}

To eastablish the monopole condensation in QCD
with the effective action has been extremely difficult
to attain. The central issue here
has been the stability of the monopole
condensation.  The earlier attempts
to prove the monopole condensation
have produced a negative result.
The effective action of QCD in the presence of
pure magnetic background calculated first
by Savvidy and subsequently by Nielsen and Olesen did
produce a non-trivial magnetic condensation \cite{savv,niel}.
Unfortunately the SNO vacuum was unstable, due to the
tachyonic modes in the functional determinant of the gluon loop.
Nielsen and Olesen have correctly conjectured
that the instability of the SNO vacuum
originates from the fact that it is not gauge invariant.
To cure this defect they have proposed the Copenhagen
vacuum \cite{niel}.

In this paper we have shown that there is much simpler
and more natural way to implement the gauge invariance, to impose
the color reflection invariance to the SNO vacuum.
The color reflection invariance clearly shows that the tachyonic
modes are the gauge artifact which are not gauge invariant.
This disqualifies them as physical states. In the absence of
the tachyons, of course, we have a stable monopole condensation
and the dimensional transmutation in QCD. This endorses
the conjecture of Nielsen and Olesen.

It is not surprising that the gauge invariance plays the crucial role
in the stability of the monopole condensation. From the beginning
the gauge invariance has been the main motivation for the confinement
in QCD. It is this gauge invariance which forbids any
colored object from the physical spectrum of QCD.
This necessitates the confinement of color.
So it is only natural that the gauge invariance
assures the stability of the monopole condensation, and thus
the confinement of color.

As we have pointed out, there are two ways to exclude the unphysical
tachyonic states in the calculation of the effective action.
One could either exclude them when one calculates the functional
determinant (\ref{gdet2}), imposing the gauge invariance properly
as we did in this paper. If one does this, the integral expression
of the effective action (\ref{eaabo}) has no infra-red
divergence and thus there is no need of any infra-red regularization.
Or one could include them at this stage, and remove them later.
If one chooses to do so, one obtains the integral expression
(\ref{eaabx}) of effective action which has a severe
infra-red divergence. In this case one must
exclude them with the infra-red regularization by causality,
not with the $\zeta$-function regularization \cite{cho3,cho4}.
The reason is that the $\zeta$-function regularization
does not remove any states included in the determinant.
This does not mean that the $\zeta$-function regularization
has any intrinsic deficiency. On the contrary the
$\zeta$-function regularization is too
honest to change the functional determinant (\ref{fdabx}).

Given the importance of the issue, however, one may need
a further varification of the monopole condensation.
One can do this with the perturbative calculation
of the imaginary part of the one-loop effective action \cite{sch,cho4}.
This is made possible, because in QCD (and in massless QED)
the imaginary part of the one-loop effective action is
of the order $g^2$. This allows us to make
a perturbative expansion of the imaginary part of the effective action.
The perturbative calculation
produces an identical result, identical to the the infra-red
regularization by causality \cite{cho4}.
This again endorses the stability of
the monopole condensation.

One might like to think that the existence of the tachyonic
eigenstates is an essential characteristic, a sacred feature, of QCD.
Indeed the tachyonic modes have been an enigma,
the Gordian knot in QCD. It was there,
and nobody knew how to resolve this puzzle. Nielsen and Olesen
treated them as a sacred feature of QCD, which made it more
mysterious. In this paper we have argued that the tachyons
are an unphysical mirage which should not have been there
in the first place. Actually it is not rare
for us to encounter tachyonic states in physics,
which appear when one does something improper
or encounters something unphysical.
Consider a spontaneously broken Abelian gauge theory
coupled to a charged scalar field. In this case tachyons
appear when one chooses a wrong vacuum, but they disappear when one
chooses the correct one. Similarly, in bosonic string
the vacuum state becomes tachyonic, but this problem disappears when
we supersymmetrize the bosonic string. In QCD we have the same
situation. Our analysis tells that the tachyonic eigenstates
appear because we have not implemented the gauge
invariance properly. With a proper implementation
of the gauge invariance, they disappear. So there is nothing
mysterious about the tachyons in QCD.

{\it To summarize, we have presented three independent arguments
which support the stability of the monopole condensation in QCD,
the gauge invariance (the color reflection invariance) of the vacuum,
the infra-red regularization of effective action by causality,
and the perturbative calculation of the imaginary part of effective
action, all of which endorse the stability of the monopole condensation.
Furthermore all these calculations have been shown to be
consistent with duality, a fundamental symmetry of
effective action in gauge theories}.
This should be enough to settle the controversy on
the stability of the monopole condensation in QCD once and for all.
With this we can conclude that the quantum fluctuation does
create a dimensional transmutation in QCD,
triggered by the monopole condensation.
This strongly implies that QCD is a theory of confinement
in which all colored objects are confined by the
dual Meissner effect.

In this paper we have considered only the pure magnetic
or pure electric background. So, to be precise,
the above result only proves the existence of a stable
monopole condensation for a pure magnetic background. To show that
this is the true vacuum of QCD, one must calculate the effective action
with an arbitrary background in the presence of the
quarks and show that the monopole condensation
remains a true minimum of the effective potential. Fortunately, one
can actually calculate the effective action with an arbitrary
constant background, and show that indeed the monopole condensation
becomes the true vacuum of $SU(2)$ QCD, at least at one-loop
level \cite{cho6}. Furthermoer, we have neglected the quarks
in this paper. We simply remark
that the quarks, just like in asymptotic freedom,
tend to destabilize the monopole condensation. In fact the stability
puts exactly the same constraint on the number of quarks as
the asymptotic freedom \cite{cho6}.

It is truly  remarkable (and surprising) that the principles of
quantum field theory allow us to demonstrate confinement
within the framework of QCD. There has been a proof of
monopole condensation in a supersymmetric QCD
\cite {witt}. Our analysis shows that one can actually establish
the existence of the confinement phase within the conventional
QCD, with the existing
principles of quantum field theory.
This should be interpreted as a
most spectacular triumph of quantum field theory itself.

{\bf Acknowledgements}

~~~We thank Professor S. Adler and Professor F. Dyson
for the fruitful discussions, and Professor C. N. Yang for
the encouragements. This work is supported in part by
the ABRL Program of Korea Science and Engineering Foundation
(R14-2003-012-01002-0) and by BK21 Project of Ministry of Education.

\end{document}